\documentstyle[12pt]{article}
\begin{document}
\pagestyle{plain}
\title{The synchrotron radiation from the Volkov
solution of the Dirac equation}
\author{Miroslav Pardy\\
Department of Physical Electronics \\
Masaryk University \\
Kotl\'{a}\v{r}sk\'{a} 2, 611 37 Brno, Czech Republic\\
e-mail:pamir@physics.muni.cz}
\date{\today}
\maketitle
\vspace{50mm}

\begin{abstract}
The aim  of this article is to show it is possible to get
the power spectrum of the synchrotron radiation from the Volkov solution
of the Dirac equation and from S-matrix. 
We also generalize the Bargmann-Michel-Telegdi equation for
the spin motion in case it involves the radiation term. This equation 
can play the crucial role for the spin motion of protons in LHC. 
The axion production in the magnetic field  described by the 
Volkov solution is discussed.

\end{abstract}
\vspace{1cm}

\section{Introduction}
The aim  of this article is to show that radiation of electrons moving
in the constant  magnetic field
can be  determined from the Volkov solution of the Dirac equation.

Around year 1947  Floyd Haber, a young staff member and technician 
in the laboratory of prof. Pollock, visually observed radiation of
electrons moving  circularly in the magnetic field of the chamber of an 
accelerator (Ternov, 1994). It occurred during adjustment of cyclic 
accelerator-synchrotron
which accelerated electrons up to 100 MeV (Elder et al.,). The radiation was
observed as a bright luminous patch on the background of the chamber
of the synchrotron. It was clearly visible in the daylight. 
In this way the ``electron light''
was experimentally revealed for the first time as the 
 radiation of relativistic electrons of large centripetal acceleration.
The radiation was identified with the Ivanenko and Pomeranchuk radiation, or
with the Schwinger radiation and 
later was called the synchrotron radiation since it was 
observed for the first time in synchrotron. The radiation was
considered as the mysterious similarly to the Roentgen mysterious x-rays. 

A number of theoretical studies on the emission of
a relativistic accelerating electron had been carried out long before 
the cited experiment.
The first steps in this line was treaded by Lienard (1898). He 
used the Larmor formula

$$P = \frac {2}{3}\frac {e^{2}}{c^{3}}\left(\frac {d{\bf v}}{dt}\right)^{2}
= \frac {2}{3}\frac {e^{2}}{m^{2}c^{3}}\left(\frac {d{\bf p}}{dt}
\right)^{2}, \eqno(1) $$
and extended it to the high-velocity particles. He also received the
total radiation of an electron following a circle of an circumference
$2\pi R$.

In modern physics, 
Schwinger (1945, 1949) used the relativistic generalization of the
Larmor formula to get the total synchrotron radiation. Schwinger also
obtained the spectrum of the synchrotron radiation from the method
which was based on the electron work on the electromagnetic field, $P = - \int ({\bf j}\cdot
{\bf E}) d{\bf x}$, where the intensity of electric field he expressed as the
subtraction of the retarded and advanced electric field of a
moving charge in a magnetic field, ${\bf E} = 
\frac{1}{2}({\bf  E}_{ret} - {\bf   E}_{adv})$, (Schwinger, 1949).

Schott in 1907 
was developing the classical theory of electromagnetic  radiation of 
electron moving in the uniform
magnetic field. His calculation was based on the Poynting vector. 
The goal of Schott was to explain the spectrum of
radiation of atoms. Of course the theory of Schott was unsuccessful
because only quantum theory is adequate to explain the emission
spectrum of atoms. On the other hand 
the activity of Schott was not meaningless because he elaborated the 
theory of radiation of charged particles moving in the electromagnetic
field. His theory appeared to be only of the academical interest for
40 years. Then, it was shown that the theory and specially his formula
has deep physical meaning and applicability.
His formula is at the present time the integral part of the 
every textbook on the electromagnetic field.

The classical derivation of the Schott formula is based on the Poynting
vector ${\bf S}$ (Sokolov et al. 1966)

$${\bf S} = \frac {c}{4\pi}{\bf E}\times{\bf H}, \eqno(2)$$
end ${\bf E}$ and ${\bf H}$ are intensities of the electromagnetic field of
an electron moving in the constant magnetic field, where the magnetic field
is in the direction of the axis $z$. In this case electron moves along
the circle with radius $R$ and the electromagnetic field is considered
in the wave zone and in a point  with the spherical coordinates $r, \theta,
\varphi $. In this case it is possible to show that the nonzero 
components of the radiated field are $ -H_{\theta } = E_{\varphi}, 
H_{\theta} = E_{\theta}$ (Sokolov et al. 1966). They are calculated
from the vector potential ${\bf A}$ which is expressed as the Fourier 
integral.

The circular classical trajectory of the electron is created by the
Lorentz force $F = (e/c)({\bf v}\times {\bf H})$. The trajectory is 
stationary when the radiative reaction is not considered. The radiative 
reaction causes the transformation of the circular trajectory to the 
spiral trajectory. In quantum mechanics, the trajectory is 
stationary when neglecting the interaction of an electron with 
the vacuum field. The interaction of an electron with the vacuum
field, causes the electron jumps from the higher energetic level to
the lower ones. In quantum electrodynamics description of the motion
of electron in a homogeneous magnetic field, the stationarity of the
trajectories is broken by including the mass operator into the wave
equation.  Then, it is possible from the mass operator to derive 
the power spectral formula (Schwinger, 1973). Different approach is
involved in the Schwinger et al. article (1976).

It was shown that the spectral formula of
the synchrotron radiation following from the quasi-classical
description of the radiation of electron moving in the magnetic field 
is given by the following expression (Berestetzkii et al., 1989; (90.24)):

$$ P(\omega) = \frac{dI}{d\omega} =
-\frac{e^{2}m^{2}\omega}{\sqrt{\pi}\varepsilon^{2}}
\left\{\int_{x}^{\infty}\Phi(\xi)d\xi + \left(\frac{2}{x} +
\frac{\hbar\omega}{\varepsilon}\chi x^{1/2}\right)\Phi'(x)\right\}, 
\eqno(3)$$
where

$$x = \left(\frac{mc^{2}}{\varepsilon}\right)^{2}
\left(\frac{\varepsilon\omega}
{\varepsilon'\omega_{0}}\right)^{2/3} \stackrel{d}{=}  
\left(\frac{\hbar\omega}{\varepsilon'\chi}\right)^{2/3};\quad  
 \varepsilon = c\sqrt{p^{2} + m^{2}c^{2}}  
\eqno(4)$$
and 

$$ \omega_{0} = \frac{v|e|H}{|\bf p|} \approx \frac{|e|H}{\varepsilon}. 
\eqno(5)$$ 
is the basic frequency of circulating electron in the magnetic field.
$\Phi(x)$ is so called the Airy function and it will be defined later.

Let us remark that in the classical limit i.e. 
with $\hbar\omega \ll \varepsilon$, or with $\varepsilon' \approx
\varepsilon$, we have $x \ll 1$
and the second term in the round brackets of (3) is very small and
equation reduces, after insertion of $\omega = \omega(x)$ from
eq. (4) to the formula expressed in the form (Landau et al., 1988; (74.13))

$$I_{l} = \frac{2e^{4}H^{2}}{\sqrt{\pi} c^{3}m^{2}}\frac{mc^{2}}{\varepsilon}
\sqrt{u}\left[-\Phi'(u) -
  \frac{u}{2}\int_{u}^{\infty}\Phi(u)du\right], \eqno(6)$$
where

$$u = l^{2/3}\left(\frac{mc^{2}}{\varepsilon}\right)^{2},\quad l = 
\frac{\omega}{\omega_{0}}, \eqno(7)$$
and $l$ is number of the harmonics of the circular trajectory of the
electron moving in the constant magnetic field. Let us also remark that formula
(6) follows also from the Schott formula if the Bessel functions of it
are replaced by the Bessel functions for harmonics with $l \gg 1$.

The emitted radiation corresponds to the energy loss of electron
moving in the magnetic field. According to Schwinger (1945), the
energy loss is 20 eV per revolution of an electron with energy
$10^{8}$ eV and radius 0.5 m.

To calculate the total radiation from the formula (3) it is necessary
to integrate over all $\omega$ from $0$ to $\varepsilon$. However it is
better to change variable using the equation (4). Using this
equation, we have $\hbar \omega = \varepsilon - \varepsilon' =
\varepsilon - \hbar\omega/(\chi x^{3/2})$ and from this equation it may
be easy to see that 

$$\hbar\omega = \left(1 - \frac{1}{1 + \chi x^{3/2}}\right) =
\frac{u}{1 + u}; \quad u = \chi x^{3/2}. \eqno(8)$$

Then, we integrate from 0 to $\infty$. After two integration per partes of
the first term in the braces of formula (3), we get the following
result (Berestetzkii et al., 1989; (90.25)):

$$ I = 
-\frac {e^{2}m^{2}\chi^{2}}{2\sqrt{\pi}\hbar^{2}}
\int_{0}^{\infty} \frac {4 + 5\chi x^{3/2} + 4\chi^{2}x^{3}}
{(1 + \chi x^{3/2})^{4}}\Phi'(x) x dx = $$

$$-\frac {e^{2}m^{2}\chi^{2}}{2\sqrt{\pi}\hbar^{2}}
\int_{0}^{\infty} \frac {4 + 5u + 4u^{2}}
{(1 + u)^{4}}\Phi'(z) z dz; \quad u = \chi z^{3/2}. \eqno(9)$$

We will show in the next text how to determine formula (9)
from the Volkov solution of the Dirac equation in the
magnetic field and from  the S-matrix method. 
From  formula (9) follows also the classical expression for the
synchrotron radiation. 

The opening angle of radiation is not small  in case of the 
nonrelativistic motion. The small opening angle is generated 
only with high energy electrons as a result of the validity of special
relativity optics. According to Winick (1987), if an electron  is
given a total energy 5 GeV, the opening angle over which synchrotron 
radiation is emitted is only 0.0001 radian, or about 0.006 degree. 
This can be regarded as a beam with the nearly parallel rays. This is
practically the same as the laser beam situation. The wave length of photons is
from zero to infinity. If we want to produce maximal energy of photons
at the very short length of photons,  it is necessary to consider
the relativistic electrons.

It it possible to consider the nonrelativistic motion of a charged 
particle in the strong electric and magnetic field. The trajectory is
a cycloid with the very small radius and it means
that the external observer sees the synchrotron radiation from the 
``straight line'', which is perpendicular to the magnetic and electric
field. The most intensive radiation is generated at the direction of
``straight line'' (Pardy, 2003b). The process is realized in the
atmosphere of the neutron stars where the magnetic field is extremely
strong.

\section{The Volkov solution of the Dirac equation 
 in the constant magnetic field}

In order to derive the classical limit of the quantum radiation
formula, we will suppose that the motion of the
Dirac electron is performed in the following four potential:

$$A_{\mu} = a_{\mu}\varphi; \quad \varphi = kx; \quad k^{2} = 0. \eqno(10)$$

From equation (10), it follows that $F_{\mu\nu} = \partial_{\mu}A_{\nu} -
\partial_{\nu}A_{\mu}  = a_{\mu}k_{\nu} - a_{\nu}k_{\mu} = const.$, which
means that electron moves in the constant electromagnetic field with the
components ${\bf E}$ and ${\bf H}$. The parameters $a$ and $k$ can be chosen
in a such a way that ${\bf E} = 0$. So the motion of electron is performed in
the constant magnetic field.

The Volkov (1935) solution of
the Dirac equation for an electron moving in a field of a plane wave
is (Berestetzkii et al., 1989; Pardy, 2003a; Pardy, 2004): 

$$\psi_{p} = \frac {u(p)}{\sqrt{2p_{0}}}
\left[1 + e \frac {(\gamma k)
(\gamma A(\varphi))}{2kp}
\right] \exp \left[(i/\hbar)S \right]\eqno(11)$$
and $S$ is an classical action of an electron moving in the 
potential $A(\varphi)$.

$$S = -px - \int_{0}^{kx}\frac {e}{(kp)}\left[(pA) - \frac {e}{2}
(A)^{2}\right]d\varphi. \eqno(12)$$

It was shown that for the potential (10) the Volkov wave function 
is (Berestetzkii et al., 1989):

$$\psi_{p} = \frac {u(p)}{\sqrt{2p_{0}}}
\left[1 + e \frac {(\gamma k)
(\gamma a)}{2kp}\varphi
\right] \exp \left[(i/\hbar)S \right]\eqno(13)$$
with

$$S = -e\frac {ap}{2kp}\varphi^{2} + e^{2}\frac {a^{2}}{6kp}\varphi^{3} -
px. \eqno(14)$$

During the following text we will suppose that we will work in the unit system
where $c = \hbar = 1$.

\section{S-matrix element for photon emission}

While the Larmor formula (1) involves explicitly the dependence of the
radiation on the derivative of the particle velocity over the
time, the quantum field theory works only with the matrix elements and power
spectrum must be determined from the correct definition of the matrix
element in case that electron is moving in potential (10). The
alternative method can be considered and it consists 
in using the quantization of the Poynting vector
(2) in the differential intensity as $dI = ({\bf
  r}\cdot {\bf S})d\Omega$. However, to our knowledge, this method was
not elaborated and  published. Similarly, the gravity radiation was not
determined from the general relativistic definition of the Poynting
vector.

The situation in the quantum physics differs from the situation in the 
classical one. The emission of photons by electron moving in the
homogeneous magnetic field is the result of the transition of electron
from the stationary state with energy $E_{a}$ to stationary state with
energy $E_{b}$, where $E_{a} >  E_{b}$. The transition between
stationary states is called spontaneous, however it is stimulated by
the interaction of an electron with the virtual electromagnetic field
of vacuum, or, in other words, by the interaction of electron with
virtual photons. So, it is necessary to find the interaction term of an
electron with vacuum photons and to solve the Dirac equation with this
term and then to determine the matrix elements of the transition.

The quantum field theory expressed as the source theory was used to
solved the synchrotron radiation by Schwinger (1973). In this language
and methodology the original action term for the spin-0 charged
particle was supplemented by the mass operator in the homogeneous 
magnetic field and it was shown that this mass operator involves as an
integral part the power spectral formula of the synchrotron radiation.
Here we use the Volkov solution of the Dirac equation and the S-matrix
approach to find the probability of emission and the intensity of the
synchrotron radiation. The method is nonperturbative because the
Volkov solution of the Dirac equation can be expressed in the
nonperturbative form. 

While the Feynman diagram approach requires renormalization
procedure and the Schwinger source methods requires contact terms  as
some form of renormalization, our method does not work with renormalization.
The wave function of an electron involves the total interaction of an
electron with magnetic field. 

The question, if the Lorentz-Dirac equation with the radiative term 
can be derived from the
S-matrix approach, or from the Feynman diagram approach is unanswered,
and to our knowledge it was not published. On the other hand the more
simple Lorentz equation for the charged particle motion in
electromagnetic field
was derived using the WKB approximation of the
Dirac equation together with the Bargmann-Michel-Telegdi equation for the
spin motion (Rafanelli and Schiller, 1964; Pardy; 1973).   

It is possible to show in the quantum field theory,  that 
the corresponding S-matrix element which describes transition from the state
$\psi_{p}$ to $\psi_{p'}$ with simultaneous emission of photon with
polarization $e'$ and four-momentum $k'^{\mu} = (k_{0}', {\bf k}') =
 (\omega', {\bf k}')$ is given
by the following expression (Berestetzkii et al., 1989), with $k' \to
-k', S \to -S $, to be in accord with the Ritus article (Ritus, 1979):

$$M = e\int d^{4}x \bar\psi_{p'} (\gamma e'^{*})\psi_{p}
\frac {e^{-ik'x}}{\sqrt{2\omega'}}, \eqno(15)$$
where $\psi_{p}$ is given by the relation

$$\psi_{p} = 
\exp {i\left\{e\frac {(ap)}{2(kp)}\varphi^{2} -  e^{2}\frac
{a^{2}}{6(kp)}\varphi^{3} +  px\right\}}
\left[1 + e\frac {(\gamma k)(\gamma a)}{2(kp)}\varphi\right]
\frac {u(p)}{\sqrt{2p_{0}}} \eqno(16)$$
and $\bar\psi_{p}$ is the the conjugated function to $\psi_{p}$, or,

$$\bar\psi_{p} = \frac {\bar u(p)}{\sqrt{2p_{0}}}
\left[1 + e\frac {(\gamma a)(\gamma k)}{2(kp)}\varphi\right]
\exp {(i)\left\{-e\frac {(ap)}{2(kp)}\varphi^{2} + e^{2}\frac
{a^{2}}{6(kp)}\varphi^{3} - px\right\}}. \eqno(17)$$

Afer insertion of eq. (16) and  (17) into eq. (15) and putting 

$$\exp\left\{i\left(\frac{\alpha\varphi^{2}}{2} -
\frac{i4\beta\varphi^{3}}{3}\right)\right\} =
\int_{-\infty}^{\infty}ds e^{is\varphi}A(s,\alpha,\beta)\eqno(18)$$
with

$$\alpha =  e\left(\frac {ap}{kp} - \frac {ap'}{kp'}\right);
\quad \beta = \frac {e^{2}a^{2}}{8}\left(\frac {1}{kp} - \frac {1}{kp'}
\right),  \eqno(19)$$
 we get (Ritus, 1979):

$$M =
e\int_{-\infty}^{\infty} \frac {ds}{\sqrt{2\omega'}}
(2\pi)^{4}\delta(p + sk - p' - k')\bar u(p')\quad \times $$

$$\left\{(\gamma e'^{*})A + e\left(\frac {(\gamma a)(\gamma k)
(\gamma e'^{*})}{2(kp')} +
\frac {(\gamma e'^{*})(\gamma k)(\gamma a)}{2(kp)}\right)i
\frac {\partial A}{\partial s} +
\frac {e^{2}a^{2}(ke'^{*})(\gamma k)}{2(kp)(kp')}
\frac {\partial^{2}A}{\partial s^{2}}\right\}u(p).\eqno(20)$$
 
It evidently follows from eq. (18), that 

$$A(s,\alpha,\beta) = \frac {1}{2\pi}\int_{-\infty}^{\infty}d\varphi
\exp\left\{i\left(\frac {\alpha\varphi^{2}}{2} -  \frac {4\beta\varphi^{3}}{3}
 - s\varphi\right)\right\}.\eqno(21)$$

The terms $i^{n}\partial^{n}A/\partial s^{n}$  are the Fourier mapping
of functions

$$\varphi^{n}\exp(i\alpha\varphi^{2}/2 - i4\beta\varphi^{3}/3). \eqno(22)$$

The matrix element $M$ is analogical to the emission of photons
generated by
the electron in the plane electromagnetic wave  $A_{\mu} = a_{\mu}\cos(kx)$
which was proved by Ritus (1979),
and the difference is in replacing the discrete $s$ by the continual quantity.
So the summation over $s$ is replaced by the integration.

Function $A(s, \alpha, \beta)$ can be expressed by the Airy function
$\Phi(y)$:

$$ A(s, \alpha, \beta) = \frac {1}{\pi}(4\beta)^{-1/3}
\exp\left\{-is\frac {\alpha}{8\beta} + i\frac {8\beta}{3}
\left(\frac {\alpha}{8\beta}\right)^{3}\right\}\Phi(y), \eqno(23)$$
where the Airy function $\Phi(y)$ is defined by the equation

$$\frac{d^{2}\Phi}{dy^{2}} - y\Phi = 0\eqno(24)$$
with the solution 

$$\Phi(y) = \frac {1}{2}\int_{-\infty}^{\infty}du\;e^{-i\left(\frac {u^{3}}{3}
 + yu\right)} = \int_0^\infty du\cos\left(\frac {u^{3}}{3} + yu\right),
 \eqno(25)$$
where in our case 

$$y = (4\beta)^{2/3}\left[\frac {s}{4\beta} -
\left(\frac {\alpha}{8\beta}\right)^{2}\right], \eqno(26)$$
where $\beta\geq 0$. Landau et al. (1988) uses the Airy function
expressed as $\Phi/\sqrt{\pi}$.

Using the formula (21) it is easy to find the differential equation
for $A(s)$:

$$sA - i\alpha A' -4\beta A'' = 0, \eqno(27)$$
where $A' = \partial A/\partial s, A'' = \partial^{2} A/\partial s^{2}$.

The evaluation of the squared matrix elements, which has physical meaning of
the probability of the radiation process, involves, as can be seen, the
double integral for which we use the identity:

$$\int_{-\infty}^{\infty}ds \int_{-\infty}^{\infty}ds' F(s)
\delta (sk + p - p' - k')\delta (s'k + p - p' - k') = $$

$$ \int_{-\infty}^{\infty}ds \int_{-\infty}^{\infty}ds'F(s)
\frac{\delta(s - s')}{\delta(0)}\delta(sk + p - p' - k') = $$

$$ \frac{VT}{(2\pi)^{4}}\int_{-\infty}^{\infty}ds 
\frac {F(s)}{\delta (0)}\delta(sk + p - p' - k'). \eqno(28)$$

So, now, we are prepared to determine the probability of the emission of
photons and we perform it in the following section.

\section{Probability of emission of photons}

Using the ingredients of the quantum field theory, we get for the probability
of the emission of one photon an electron in unit volume per unit
time (Ritus, 1979):

$$\sum_{r,r'}\frac {|M|^{2}}{VT} = \frac {(2\pi)^{5}e^{2}}{\delta(0)}
\int_{-\infty}^{\infty}\frac {ds}{2p_{0}p'_{0}\omega'}
\delta(sk + p - p' - k') \quad \times $$

$$\left\{|pe''^{*}A - ieae''^{*} A'| -
2\beta(kk')(|A'|^2  + {\rm Re}AA''^{*})\right\}, \eqno(29)$$
where

$$e''_{\alpha} = e_{\alpha}' - k_{\alpha}'(ke')/(kk');
\quad A' = \partial A/ \partial s; \quad  A'' = \partial^{2} A/\partial
s^{2} . 
\eqno(30)$$

After summation of eq. (29) over directions of polarization $e'$ and using
differential equation (27), $\sum |M|^{2}/VT$ will be expressed only
by means of $|A|^{2}$ and $|A'|^{2} + {\rm Re}AA''^{*}$, which can be
expressed using eq. (23) in the following way:

$$ |A|^{2} = \frac {\Phi^{2}(y)}{\pi^{2}(4\beta)^{2/3}};\quad
|A'|^{2} + {\rm Re AA''^{*}} = \frac{y\Phi^{2}(y) + \Phi'^{2}(y)}{
{\pi^{2}(4\beta)^{2/3}}}. \eqno(31)$$

Then, with 

$$
x = \frac {ea}{m}, \quad \chi = - \frac{kp}{m^{2}}x, \quad 
\chi' = - \frac{kp'}{m^{2}}x ,\quad \kappa = -\frac{kk'}{m^{2}}x, \eqno(32)$$
we have

$$\sum_{r,r'}\frac {|M|^{2}}{VT} = \frac {2 e^{2}m^{2}}
{\delta(0)x^{2}p_{0}p_{0}'k_{0}'}
\int_{-\infty}^{\infty}ds\left(\frac {2\chi\chi'}{\kappa}\right)^{2/3}
\delta(sk + p - p' - k') \quad \times $$

$$\left\{-\Phi^{2}(y) + \left(\frac {2\chi\chi'}{\kappa}\right)
\left(1 + \frac {\kappa^{2}}{2\chi\chi'}\right)^{2/3}
\left[y\Phi^{2}(y) + \Phi'^{2}(y)\right]\right\}. \eqno(33)$$

In order to obtain the probability of the emission of photon by electron, it i
necessary to integrate equation (33) over the final states
$d^{3}p'd^{3}k'(2\pi)^{-6}$ of the electron and photon and the result divide
by 1/2  and to average over polarizations of the initial electron.
Integration over ${\bf p}'$ eliminates space $\delta$-function and the time
$\delta$-function can be transformed into the explicit Lorentz invariant form
as follows:

$$\delta(sk + p - p' - k') \frac {d^{3}p'}{p_{0}'}\quad \rightarrow \quad
\frac{\delta(sk_{0} + p_{0} - p_{0}' - k_{0}')}{p_{0}'} = - \frac
{\delta(s - \tilde s)}{kp'}; \quad \tilde s = \frac {k'p'}{kp}\eqno(34)$$
with the use of the relation

$$p_{0}' = \sqrt{m^{2} + (s{\bf k} + {\bf p} - {\bf k}')^{2}}. \eqno(35)$$

Using the equation (34) and after integration over $s$, we get the
differential probability of the emission of photon per unit time:

$$dW = \frac {e^{2}c}{4\pi^{3}x\delta(0)\chi'}\left(\frac
{2\chi}{u}\right)^{2/3}\quad \times $$

$$\left[-\Phi^{2}(y) + \left(\frac {2\chi}{u}\right)^{2/3}
\left(1 + \frac {u^{2}}{2(1 + u)}\right)
(y\Phi^{2}(y) + \Phi'^{2}(y))\right]\frac {d^{3}k'}{k_{0}'}, \eqno(36)$$
where

$$u = \frac {\kappa}{\chi'} = \frac{kk'}{kp'}, \quad c = \frac
{1}{p_{0}}. \eqno(37)$$

The equation (36) is evidently relativistic and gauge invariant. The further
properties are as follows. It does not depend on $k_{1}'$, which is the
component of the photon momentum along the electric field ${\bf E}$.
It means it does not depend on $\varrho$. We use further the transform

$$\frac {d^{3}k'}{k_{0}'} = 
\frac{x m^{2}\chi' u}{\chi(1 + u)^{2}}d\varrho d\tau du.\eqno(38)$$
with the obligate relation (Ritus, 1979):

$$\int_{-\infty}^{\infty}d\varrho = \delta(0).\eqno(39)$$

After integration of (36) over $\varrho$ and with regard to (39), we get
the probability of emission of photons in variables $u, \tau$ without
dependence of the localization of the emission the following formula:

$$dW = \frac {e^{2}m^{2}c}{2\pi^{3}(1 + u)^{2}}\left(\frac
{u}{2\chi}\right)^{1/3}\quad \times $$

$$\left[-\Phi^{2}(y) + \left(\frac {2\chi}{u}\right)^{2/3}
\left(1 + \frac {u^{2}}{2(1 + u)}\right)
(y\Phi^{2}(y) + \Phi'^{2}(y))\right]dud\tau. \eqno(40)$$

The probability (40) has a dimension of ${\rm cm}^{-3}{\rm s}^{-1}$.

Formula (40) describes the dependence of the distribution of probability
on two variables $u, \tau$. If we use equation

$$y\Phi^{2}(y) + \Phi'^{2}(y) = \frac {1}{2}\frac {d^{2}}{dy^{2}}\Phi^{2}(y)
\eqno(41)$$
and the transformation $t = a \tau^{2};\quad a = \left(\frac
  {u}{2\chi}\right)^{2/3}, d\tau = \frac{dt}{2\sqrt{at}}$, then, 
we get with $y = a + t$ the following result

$$\frac{dW}{du} = \frac {e^{2}m^{2}c}{2\pi^{3}(1 + u)^{2}}\quad \times $$

$$\left\{-1 + \left(\frac {2\chi}{u}\right)^{2/3}
\left[1 + \frac {u^{2}}{2(1 + u)}\right]
\frac {1}{2}\frac {d^{2}}{da^{2}}\right\}
\int_{0}^{\infty}\frac {dt}{\sqrt{t}}\Phi^{2}(a + t); \quad
a = \left(\frac {u}{2\chi}\right)^{2/3}. \eqno(42)$$

Now, let us use the integral transformation (Aspnes, 1966)

$$\int_{0}^{\infty}\frac {dt}{\sqrt{t}}\Phi^{2}(a + t) = \frac {\pi}{2}
\int_{2^{2/3}a}^{\infty}dy \Phi(y). \eqno(43)$$

Then, we get from the formula (42)

$$\frac{dW(\chi, u)}{du} = - \frac {e^{2}m^{2}c}{4\pi^{2}(1 + u)^{2}}
\quad \times $$

$$\left\{\int_{z}^{\infty}dy\Phi(y) + \frac {2}{z}
\left[1 + \frac {u^{2}}{2(1 + u)}\right]
\Phi'(z)\right\}; \quad
z = \left(\frac {u}{\chi}\right)^{2/3}. \eqno(44)$$

The last formula can be written easily for small and big $u$ as follows:

$$\frac{dW}{du} = - \Phi'(0)\frac {e^{2}m^{2}c}{2\pi^{2}}
\left(\frac {\chi}{u}\right)^{2/3} ; \quad u \ll 1, \quad \chi; \eqno(45)$$
where 

$$\Phi'(0) = - \frac{1}{3^{1/3}}\int_{0}^{\infty}x^{-1/3}\sin x\; dx = 
 - \frac{1}{3^{1/3}}
\int_{0}^{\infty} x^{\mu -1}\sin(ax)\;dx\big |_{\mu = 2/3; a = 1} = $$
 
$$- \frac{1}{3^{1/3}}\frac{\Gamma(\mu)}{a^{\mu}}
\sin\left(\frac{\mu\pi}{2}\right)
\Big|_{\mu = 2/3; a = 1} =  - 3^{1/6}\frac {\Gamma(2/3)}{2}, \eqno(46)$$

$$\frac{dW}{du} =  \frac {e^{2}m^{2}c}{8\pi^{3/2}u^{3/2}}\sqrt{\chi}
\exp{\left(\frac {-2u}{3\chi}\right)}
; \quad u \gg 1, \quad \chi . \eqno(47)$$

If we integrate the formula (44) over $u$ and using the per partes method in
the first term, we get the following formula:

$$W(\chi) = - \frac {e^{2}m^{2}c}{8\pi^{2}}\chi
\int_{0}^{\infty}dz
\frac {5 + 7u + 5u^{2}}{\sqrt{z}(1 + u)^{3}} \Phi'(z)
; \quad u = \chi z^{3/2} . \eqno(48)$$

The last formula was derived for the first time by Goldman 
(1964a,  1964b) by the
different way. This formula can be expressed approximately for small and big
$\chi$ as follows (Ritus, 1979):

$$W(\chi) = - \frac {5e^{2}m^{2}c}{8\sqrt{3}\pi}\chi
\left(1 - \frac {8\sqrt{3}}{15}\chi + ...\right); \quad  \chi \ll 1,
\eqno(49)$$

$$W(\chi) = - \frac {7\Gamma(2/3) e^{2}m^{2}c}{54\pi}(3\chi)^{2/3}
\left(1 - \frac {45}{28\Gamma(2/3)}(3\chi)^{-2/3} + ...\right); \quad  \chi
\gg 1 . \eqno(50)$$

\section{Intensity of radiation}

Ritus (1979) proved that the probability of radiation and the intensity of
radiation differs only by the specific term beyond the integral
function. So, using the Ritus proof and with regard to eq. (8) we see that  
the intensity of radiation can be obtained from formula (40) putting
$c/p_{0}\quad\rightarrow \quad 1$ and by the multiplication by the term
$u(1+u)^{-1}$. We get:

$$dI = - \frac {e^{2}m^{2}}{2\pi^{3}}\frac {u}{(1 + u)^{3}}\quad \times $$

$$\left(\frac {u}{2\chi}\right)^{1/3}
\left\{-\Phi^{2}(y) + \left(\frac {2\chi}{u}\right)^{2/3}
\left(1 + \frac {u^{2}}{2(1 + u)}\right)
(y\Phi^{2}(y) + \Phi'^{2}(y))\right\}du d\tau. \eqno(51)$$

Then the $u$-distribution over intensity is of the form:

$$\frac{dI}{du} = -\frac {e^{2}m^{2}}{4\pi^{2}}\frac {u}{(1 + u)^{3}}
\left\{\int_{z}^{\infty}dy\Phi(y) + \frac {2}{z}
\left(1 + \frac {u^{2}}{2(1 + u)}\right)
\Phi'(z))\right\};\quad z = \left(\frac {u}{\chi}\right)^{2/3}.
\eqno(52)$$

This formula is a quantum generalization of the classical expression 
for the spectral distribution of radiation of an ultrarelativistic
charged particle in a magnetic field (Landau et al., 1988; (74.13)).
 
After integration of (52) over $u$ and using the per partes method in the
first term, we get the formula of the total radiation of photons by electron
in the constant magnetic field. The formula is as follows (Ritus, 1979):

$$I = -\frac {e^{2}m^{2}}{2\sqrt{\pi}\hbar^{2}}\chi^{2}
\int_{0}^{\infty}dz z \frac {4 + 5u + 4u^{2}}{2(1 + u)^{4}}
\Phi'(z);  \quad u = \chi z^{3/2}. \eqno(53)$$

This formula can be transformed with $u = \chi x^{3/2}$ to
the following one:

$$I = -\frac {e^{2}m^{2}\chi^{2}}{2\sqrt{\pi}\hbar^{2}}
\int_{0}^{\infty}\frac {4 + 5\chi x^{3/2} + 4\chi^{2}x^{3}}
{(1 + \chi x^{3/2})^{4}}\Phi'(x) x dx. \eqno(54)$$

The formula (54) is equivalent with the formula (3) and the formula 
(52) is identical with the formula (3). There is no doubt that the
Schott formula can be derived by the formalism used in this text.

\section{Discussion}

We have seen how to get the quantum description of the synchrotron
radiation from the Volkov solution of the Dirac equation and from the
formalism of the relativistic quantum theory of radiation. At the same
time we have shown that the quantum synchrotron radiation leads to the
classical synchrotron radiation in the classical limit.

The synchrotron radiation evidently influences the motion of
the electron in accelerators. The corresponding equation which describes
the classical motion is so called the Lorentz-Dirac equation, which
differs from the the so called Lorentz equation only by the additional
term which describes the radiative corrections. The equation with the
radiative term is as follows (Landau et al., 1988):

$$m \frac{dv_{\mu}}{ds} = \frac{e}{c}F_{\mu\nu}v^{\nu} +
g_{\mu},\eqno(55)$$
where the radiative term was derived by Landau et al. in the form 
(Landau et al., 1988)

$$g_{\mu} = \frac{2e^{3}}{3mc^{3}}\frac{\partial F_{\mu\nu}}
{\partial x^{\alpha}}v^{\nu}v^{\alpha} - 
\frac{2e^{4}}{3m^{2}c^{5}} F_{\mu\alpha}F^{\beta\alpha}v_{\beta} + 
\frac{2e^{4}}{3m^{2}c^{5}} \left(F_{\alpha\beta}v^{\beta}\right) 
\left(F^{\alpha\gamma}v_{\gamma}\right)v_{\mu}. \eqno(56)$$

Bargmann, Michel and Telegdi (Berestetzkii, 1989;) derived so called 
BMT equation for motion of spin in the electromagnetic field,  in the form 

$$\frac{da_{\mu}}{ds} = 2\mu F_{\mu\nu}a^{\nu} -2\mu'v_{\mu}
F^{\nu\lambda}v_{\nu}a_{\lambda},\eqno(57)$$
where $a_{\mu}$ is so called axial vector describing the classical
spin. It was shown by Rafanelli and Schiller (1964), (Pardy, 1973) 
that this equation
can be derived from the classical limit, i.e. from the WKB solution 
of the Dirac equation with the anomalous magnetic moment. 

It is meaningful to consider the BMT equation with the radiative
corrections to express the influence of the synchrotron radiation on
the motion of spin. To our knowledge such equation, the generalized
BMT equation, was not published
and we here present the conjecture of the form of such equation. The
equation is of the following form:

$$\frac{da_{\mu}}{ds} = 2\mu F_{\mu\nu}a^{\nu} -2\mu'v_{\mu}
F^{\nu\lambda}v_{\nu}a_{\lambda} + g_{(axial)\mu},\eqno(58)$$
where the term $g_{(axial)\mu}$ is generated as the "axialization"' of
the radiation term $g_{\mu}$. Or, 

$$g_{\mu} = \frac{2e^{3}}{3mc^{3}}\frac{\partial F_{\mu\nu}}
{\partial x^{\alpha}}v^{\nu}a^{\alpha} - 
\frac{2e^{4}}{3m^{2}c^{5}} F_{\mu\alpha}F^{\beta\alpha}a_{\beta} + 
\frac{2e^{4}}{3m^{2}c^{5}} \left(F_{\alpha\beta}v^{\beta}\right) 
\left(F^{\alpha\gamma}v_{\gamma}\right)a_{\mu}. \eqno(59)$$

We are aware that the axialization is not unambiguous and it is
evident, that it can be submitted
for theoretical investigation. The future physics will give the answer if
the equation corresponds to physical reality. Such equation will
have fundamental meaning for the work of LHC where the synchrotron
radiation influences the spin motion of protons in LHC.

The formalism used in case of the synchrotron radiation 
 can be also applied in the situation
where the axion is produced in the magnetic field. Axion was
introduced by Peccei and Quinn (1977) as the pseudoscalar particle.
It was introduced as the
logical necessity of the correct physical theory and it means that
there is the great probability that axions will be detected for instance
during the experiments on LHC. 

One
of the corresponding Lagrangian  describing the interaction of the
axion field $a$ with the electron field $\psi$ is as follows
(Skobelev, 1997):

$${\cal L} = -ic\left(\frac{m_{a}}{f}\right)a(\bar \psi\gamma^{5}\psi),
\eqno(60)$$
where $f$ is related to the coupling constant.

According to Skobelev (1997) the intensity if emission of axions by
the electron moving in the constant electromagnetic magnetic field
can be approximated by two formulas which follows from the general
theory.

$$I_{a} = \frac{g^{2}m^{2}}{\pi}\chi^{4}; \quad g = \frac{c m}{f}; \quad 
\chi \ll 1; \quad \frac{m_{a}}{m} \ll \chi ,\eqno(61)$$
and/or

$$I_{a}  = \frac{7\Gamma(2/3)g^{2}m^{2}}{2\pi 3^{13/3}}\chi^{2/3};
\quad \chi \gg 1; \quad \frac{m_{a}}{m} \ll \chi.  \eqno(62)$$

Axion is used also to explain the absence
of the electrical dipole moment of the neutron. Axion is chargeless,
spinless and interact with the ordinary matter only very weakly. If it
is not confined, then the following decay equation is valid:

$$n \rightarrow e + p + \bar\nu_{e} + a,   \eqno(63)$$
which can be verified in experiment as the proof of the existence of
axion in a sense that axion can decay into neutrinos as follows

$$a  \rightarrow \nu_{e} + \bar\nu_{e}. \eqno(64)$$

On the other hand if we use the plasma of particles $e, p, \bar\nu_{e},
a$, then the inversion equation to (88) is valid:

$$ e + p + \bar\nu_{e} + a   \rightarrow  n \eqno(65)$$
and this equation can be also used as the proof of the existence of
axion. If we prepare the same plasma without axions, then no neutron
will be generated. It seems that these simple experiments can be
considered as a crucial ones for the proof of the existence of axions.  

The decay of neutron and axion can be considered and calculated in 
the electromagnetic field as was shown by Skobelev (1997; 1999). 

Khalilov et al. (1995) calculated production of the of $W^{-}$ and
$Z^{0}$ bosons by electron in the intense electromagnetic field. For
the first process they used the following matrix element

$$M_{e\to W} = -i\frac{g}{2\sqrt{2}}\int d^{4}x \bar\psi_{\nu}(1 -
\gamma^{5})\gamma^{\mu}\psi_{e}\phi_{\mu}, \eqno(66)$$
where $\psi_{\nu},\psi_{e},\phi_{\mu}$ are wave functions of neutrino,
electron and $W$-boson.

In case for the production of the $Z$-boson Khalilov et al. used the
following matrix element

$$M_{e\to Z} = {\tilde g}\int d^{4}x \bar\psi_{e}\gamma^{\lambda}
(g_{V} + g_{A}\gamma^{5})\psi_{e}Z_{\lambda}. \eqno(67)$$

It has been calculated the probability of creation and the
total cross-section to every process.

It is evident that all interaction of particle
physics occurring in the accelerators and LHC can be immersed into
the intense electromagnetic field of laser, or laser pulse or magnetic
field. The theoretical investigation can then be performed 
using the Volkov solution and the S-matrix method. This will obviously
become the integral part of the future physics of elementary particles.
  
\vspace{7mm}

\noindent
{\bf References}

\vspace{7mm}

\noindent 
Aspnes, D. E., (1966). Electric-field effects on optical absorption near
thresholds in solids, {\it Phys. Rev.} {\bf 147}, 554.\\[2mm]
Berestetzkii, V. B., Lifshitz, E. M. and Pitaevskii L. P., (1989).
{\it Quantum electrodynamics} (Moscow, Nauka). \\[2mm]
Elder, F. R., Langmuir, R. V. and Pollock,  H.  C. (1948). 
Radiation from Electrons Accelerated in a Synchrotron,  {\it
Phys. Rev}. {\bf 74}, 52. \\[2mm]
Goldman, I. I., (1964a). Intensity effects in Compton scattering, 
ZhETF, 46, 1412. \\[2mm]
Goldman, I. I., (1964b). Intensity effects in Compton scattering,  {\it
  Phys. Lett.}, {\bf 8}(2), 103.\\[2mm]
Khalilov, V. P. and Dorofeev, O. F., (1995). Production of $W^{-}$ and
$Z^{0}$ bosons by electron in the intense electromagnetic field,
Nuclear Physics, {\bf 58}, No. 10, 1850. (in Russian).\\[2mm] 
Landau, L. D.  and Lifshitz, E. M., (1988). The Classical
Theory of Fields, 7th ed., (Moscow, Nauka) (in Russian). \\[2mm]
Lienard A., L' Eclairage Elec., 16, (1898) 5.\\[2mm]
Pardy,  M., (1973).
Classical motions of spin 1/2 particles with zero anomalous
magnetic moment, {\it Acta Phys. Slovaca} {\bf 23}, No. 1, 5. \\[2mm]
Pardy, M., (2003a). Electron in the ultrashort laser pulse,
{\it International Journal of Theoretical Physics}, {42}(1) 99.\\[2mm]
Pardy, M., (2003b). Theory of the magnetronic laser, 
e-print  physics/0306024.\\[2mm]
Pardy, M., (2004). Massive photons and the Volkov solution,
{\it International Journal of Theoretical Physics}, {43}(1) 127.\\[2mm]
Peccei, R. D. and Quinn, H. R., (1977). CP conservation in the presence
of instantons,  {\it Phys. Rev. Lett.} 
{\bf 38}, 1440. \\[2mm]
Rafanelli, K, and Schiller, R., (1964). Classical motion of
spin-1/2 particles, {\it Phys. Rev.} {\bf 135}, No. 1 B, B279.\\[2mm]
Ritus, V. I., (1979).  The quantum effects of the interaction of elementary
particles with the intense electromagnetic field, Trudy FIAN {\bf 111},
pp. 5-151. (in Russian). \\[2mm]
Schwinger, J., (1945). On Radiation by Electrons in Betatron, LBNL-39088, CBP
Note-179, UC-414.\\[2mm]
Schwinger, J., (1949).  On the Classical Radiation of Accelerated
Electrons, {\it Phys. Rev.} {\bf 75}, No. 12, 1912.\\[2mm]
Schwinger, J., (1973). Classical radiation of accelerated electrons
II., a quantum viewpoint, {\it Phys.Rev.D} {\bf 7}, 1696. \\[2mm]
Schwinger, J.,  Tsai, W. Y.  and  Erber,  T., (1976).
Classical and quantum theory
of synergic synchrotron-\v Cerenkov radiation,
{\it Annals of Physics (New York)}, {96}(2) 303. \\[2mm]
Skobelev, V. V.,  (1997). The synchrotron and the annihilation channels
of the production of an axion in an  external electromagnetic field,
{\it ZhETF }{\bf 112}, No. 1(7), 25. (in Russian). \\[2mm]
Skobelev, V. V., (1999). Neutrino decay  of an axion in an external 
electromagnetic field, {\it Russian Physics Journal}, {\bf  42},
No. 2, 136. \\[2mm]
Sokolov, A. A., Ternov, I. M.,  Bagrov, V. G. and Rzaev R. A. (1966). 
The quantum theory of radiation of relativistic electron moving in the 
constant homogenous magnetic field. In:  Sokolov, A. A., Ternov, I. M., 
Editors, (1966). The synchrotron radiation, (Moscow, Nauka). (in
Russian). \\[2mm]
Ternov, I. M., (1994). Synchrotron radiation, {\it Uspekhi
fizicheskih nauk}, {\bf 164}(4), 429. \\[2mm]
Volkov, D. M.,  $\ddot{\rm U}$ber eine Klasse von L$\ddot{\rm o}$sungen
der Diracschen Gleichung, {\it Zeitschrift f$\ddot{\it u}$r Physik},
{\bf 94} (1935) 250.\\[2mm]
Winick, H., (1987). Synchrotron radiation, {\it  Scientific American}
{\bf 257}, November, 72.

\end{document}